\documentclass{aa}
\usepackage{times}
\usepackage{natbib}
\usepackage[dvips]{graphicx}
\usepackage{psfig,longtable,lscape}
\usepackage{psfig}
\usepackage{rotating}
\bibpunct{(}{)}{;}{a}{}{,}

\def\1E{1E1207.4-5209}
\def\XMM{{\em XMM--Newton}}
\def\EPIC{{\em EPIC}}
\def\MOS{{\em MOS}}
\def\pn{{\em pn}}
\def\ltsima{$\; \buildrel < \over \sim \;$}
\def\simlt{\lower.5ex\hbox{\ltsima}}
\def\gtsima{$\; \buildrel > \over \sim \;$}
\def\simgt{\lower.5ex\hbox{\gtsima}}

\begin{document}

\title {A Deep \XMM~Serendipitous Survey of a middle--latitude area}

\author{G. Novara\inst{1}, N. La Palombara\inst{1}, N. Carangelo\inst{1,2}, A. De Luca\inst{1}, P. A. Caraveo\inst{1,3}, R. P. Mignani\inst{4}, G. F. Bignami\inst{3,5,6}}

\institute {INAF/IASF-Milano, via Bassini 15, I--20133 Milano (I)
         \and Universit\`a di Milano--Bicocca, Piazza della Scienza 3, I--20126 Milano (I)
         \and Centre d'\'Etude Spatiale des Rayonnements (CESR), CNRS--UPS, 9 Avenue du colonel Roche, F--31028 Toulouse (F)
         \and European Southern Observatory, Karl Schwarzschild Strasse 2, D--85740 Garching (D)
         \and Universit\`a di Pavia, Dipartimento di Fisica Teorica e Nucleare, Via Ugo Bassi 6, I--27100 Pavia (I)
	 \and INFN - Sezione di Pavia, Via Ugo Bassi 6, I--27100 Pavia (I)
}

\offprints{Giovanni Novara, novara@mi.iasf.cnr.it}

\date{Received / Accepted}

\authorrunning{G. Novara et al.}

\titlerunning{A Deep \XMM~Serendipitous Survey}

\abstract{The radio quiet neutron star \1E has been the target of a 260 ks \XMM~observation, which
yielded, as a by product, an harvest of about 200 serendipitous X--ray sources above a limiting flux
of $2 \times 10^{-15}$ erg cm$^{-2}$ sec$^{-1}$, in the 0.3-8 keV energy range.  In view of the
intermediate latitude of our field ($b\simeq 10^{\circ}$), it comes as no surprise that the
log$N$--log$S$ distribution of our serendipitous sources is different from those measured either in
the Galactic Plane or at high galactic latitudes.  Here we shall concentrate on the analysis of the
brightest sources in our sample, which unveiled a previously unknown Seyfert--2 galaxy.

  }

\maketitle

\section{Introduction}

The radio quiet neutron star \1E has been the target of a 260 ks \XMM~observation
\citep{DeLuca2004}.  Such an observation ranges amongst the longest ever performed by \XMM~and, as
of today, is certainly the longest one at intermediate galactic latitude (i.e.  $|b|\simeq
10^{\circ}$).

The deepest X--ray surveys performed, such as the {\em Chandra Deep Field South}
\citep{Giacconi2001ApJ551,Rosati2002,Giacconi2002ApJS139} and {\em North} \citep{Brandt2001}, as
well as the {\em XMM Lockman Hole} survey \citep{Hasinger2001,Mainieri2002}, encompass only high
latitude regions, where serendipitous surveys were also performed \citep{Barcons2002,
DellaCeca2004}.  On the other hand, X--ray studies of the galactic population have been performed
only along the Galactic Plane:  shallow, wide--field surveys were obtained by {\em ROSAT}
\citep{Motch1998,Morley2001} and \XMM~\citep{Hands2004}, while deep, pencil--beam observations of
the Galactic Center have been performed by {\em CHANDRA} \citep{Muno2003}.

Thus, our long observation at intermediate latitude appears to be well suited to address important
issues such as the ratio between galactic and extragalactic contributors.  The combination of the
low flux limit, the wide energy band and the relatively low galactic latitude of this field has the
potential for an extremely interesting mix of source types.  Owing to the high--energy sensitivity
of \EPIC, we expect to see through the galactic disk to the distant population of QSOs, AGNs and
normal galaxies.  On top of such an extragalactic population, however, our field also samples in
great depth our Galaxy.  Here again the wide energy range allows to sample both hard and soft
sources, e.g.  population of X--ray binaries and normal stars.

Characterization of the sources' X-ray spectra, as well as the search for their optical
counterparts, are the classical tools to identify, either individually or on a statistical ground,
our sample of relatively faint sources.  Given the range of $f_{x}/f_{opt}$ values characteristic
for the known classes of X--ray sources \citep{Krautter1999}, we ought to reach $V \simeq 25$ in the
optical follow--up in order to be able to identify the majority of our serendipitous sources.  Thus,
although useful for a first filtering, Digital Sky Surveys are not deep enough for our purpose and
they do not provide an adequate color coverage.

A proposal for the complete optical coverage of the \EPIC~field at the 2.2 m ESO telescope has
already been accepted.  Waiting for its results, here we outline our detection technique as well as
the global results of such an analysis.  Next we shall focus on the analysis of the brightest
sources leading to the spectral characterization of a serendipitously discovered Seyfert--2 galaxy.

\section{X--ray analysis}

\subsection{Observations and data processing}

\begin{figure}[t]
\includegraphics[width=9cm]{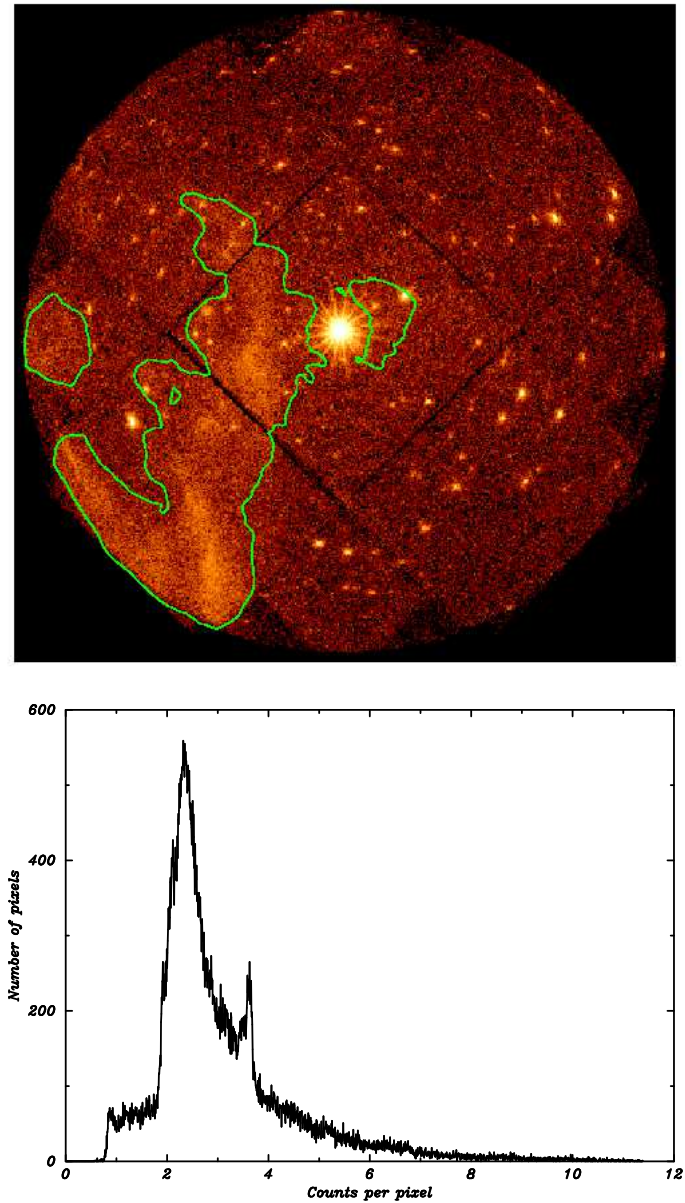}
\includegraphics[height=9cm,angle=-90]{istogramma_bkg.ps}
\begin{normalsize}
\caption{{\em Top}:  \EPIC~\MOS~image (in the energy range 0.3--8 keV) of the field of \1E.  {\em
Bottom}:  Histogram of the count number per pixel in the background map, in the energy range 0.5--2
keV.  The sky region corresponding to the tail of the distribution, at values higher than 4, is
enclosed by a green line:  it is clearly associated to the area of diffuse emission.}\label{dr}
\end{normalsize}
\vspace{-0.5 truecm}
\end{figure}

\XMM~observed \1E~during revolutions 486 and 487, which resulted in two different pointings
separated by $\sim$ 13 h.  All the three \EPIC~focal plane cameras \citep{Turner2001,Struder2001}
were active during both pointings:  the two \MOS~cameras were operated in {\em Full Frame} mode, in
order to cover the whole {\em field--of--view} of 30 arcmin; the \pn~camera was operated in {\em
Small Window} mode, where only the on--target CCD is read--out, in order to time tag the photons and
provide accurate arrival time information.  While the \pn~data have been used by \citet{Bignami2003}
and \citet{DeLuca2004} to study the radio--quiet neutron--star \1E, here we shall use the \MOS~data
to assess the population of {\em serendipitous} sources emerging from this long galactic
observation.  For both cameras the {\em thin} filter was used.

The event files were processed with the version 5.4.1 of the \XMM~{\em Science Analysis Software}
({\em SAS}).  After the standard processing pipeline, we looked for periods of high instrument
background, due to flares of protons with energies less than a few hundred keV hitting the detector
surface.  Such soft proton flares enhance the background and the corresponding time intervals have
to be rejected, reducing, accordingly, the good integration time.  In our case, the effective
observing time was $\sim$ 230 ks over a total observing time of 260 ks.
\vspace{-0.25 truecm}

\subsection{Source detection}

In order to maximize the {\em signal--to--noise} ratio ({\em S/N}) of our serendipitous sources and
to reach lower flux limits, we `merged' the data of the two cameras and of the two pointings.  We
performed the source detection in several energy ranges; first, we considered the two `classical',
coarse energy ranges 0.5--2 and 2--10 keV; then, we considered a finer energy division between 0.3
and 8 keV (since above 8 keV the instrument effective area decreases rapidly).  For each energy band
we generated the field image, the corresponding exposure map (to account for the mirror vignetting)
and the relevant background map.  The background maps were also corrected {\em pixel by pixel}, as
described in \citet{Baldi2002}, in order to reproduce the local variations.

We had also to take into account that the \XMM~image includes a region of diffuse emission
characterized by more than 4 events/pixel (Fig.~\ref{dr}), due to the SNR G296.5+10.0.  Therefore,
we performed the source detection with an `ad hoc' tuning of the parameters inside and outside the
SNR area.

\begin{figure}[!t]
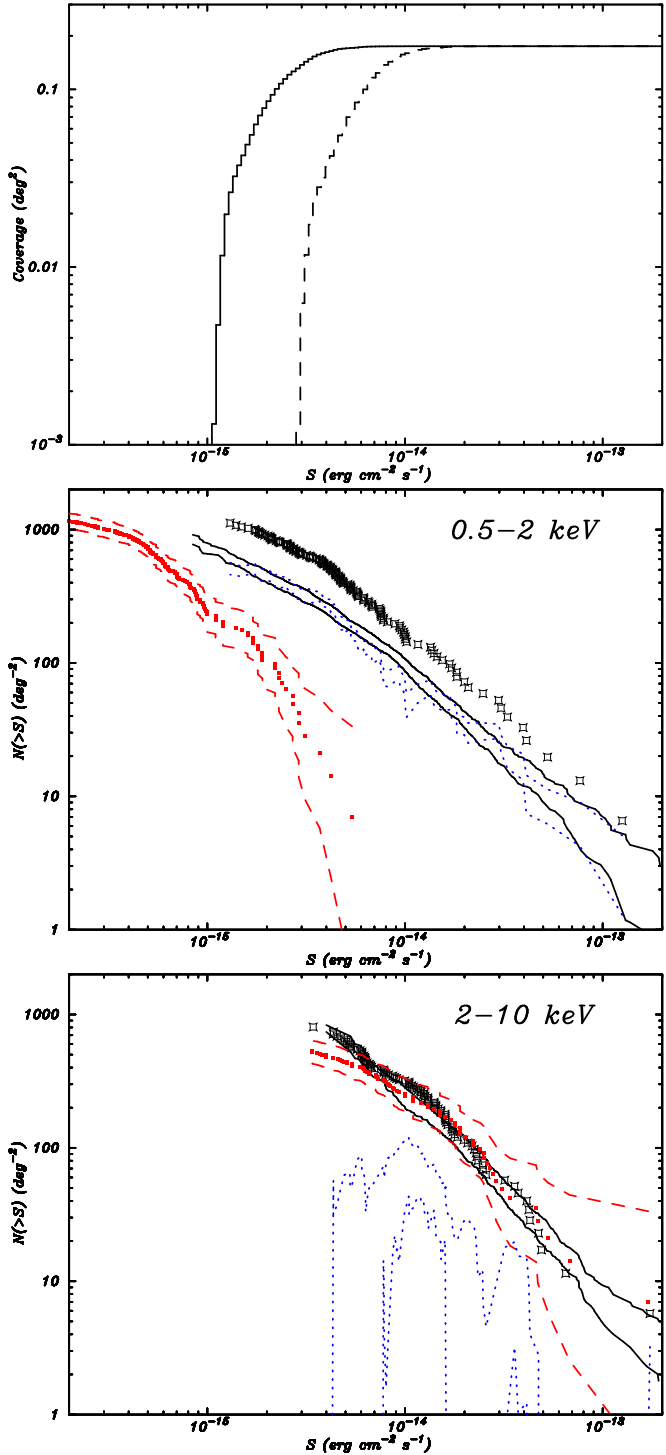

\includegraphics[height=8.75cm,angle=-90]{histo_p3_final_corr.ps}
\includegraphics[height=8.75cm,angle=-90]{all_soft_corr.ps}
\includegraphics[height=8.75cm,angle=-90]{all_hard_new.ps}
\begin{normalsize}
\caption{Sky coverage of the performed observation ({\em top}), in the energy ranges 0.5--2 keV
({\em solid line}) and 2--10 keV ({\em dashed line}), and log$N$--log$S$ distribution of the
detected sources {\em (black open squares)} in the energy ranges 0.5--2 keV ({\em middle}) and 2--10
keV ({\em bottom}).  The black solid lines trace the upper and lower limits obtained by
\citet{Baldi2002} in the same energy ranges but at higher galactic latitudes; the blue dotted lines
are the difference between our data and the Baldi et al.  ones.  The red filled squares and the red
dashed lines represent, respectively, the distributions and the limits measured by {\em CHANDRA} in
the galactic plane \citep{Ebisawa2005}.}\label{logNlogS}
\end{normalsize}
\vspace{-0.5 truecm}
\end{figure}

The source detection was based on the standard {\em maximum detection likelihood} criterium:  for
each source and each energy range we calculated a detection likelihood {\em L=-lnP}, where {\em P}
is the probability that the source counts originate from a background fluctuation.  We considered a
threshold value $L_{th}$=8.5, corresponding to a probability $P_{th}=2\cdot 10^{-4}$.  The actual
sky coverage in the various energy ranges was calculated as described in \citet{Baldi2002}:  in
Fig.~\ref{logNlogS} we show such a coverage for the two coarse energy ranges.

The  number of  spurious  detections in  each  energy range,  obtained
multiplying  $P$ times  the  number of  independent (not  overlapping)
detection cells,  is negligible.   Indeed, in our  detection procedure
the  area covered by  each cell  ranges between  0.16 and  0.35 square
arcminutes  (following  the   position  dependent  {\em  Point  Spread
Function}    size)    so   that    the    $\sim$700   square    arcmin
\EPIC~field--of--view  contains,   at  most,  5$\times10^3$  detection
cells.   Thus the  number  of spurious  detection  is $P_{th}\times  N
\leq 1$.  Since we  performed  the source  detection  in 6  independent
energy bands,  we expect the total  number of  spurious detected
sources to be at most 6.  Selecting  all the sources with  $L>8.5$ in at
least one of our energy  ranges and matching those detected in several
energy  intervals we found  a total  of 196  sources (with  a position
accuracy of  $\sim$ 5''),  35 inside the  area covered by  the diffuse
emission and 161 outside it.   We detected 135 sources between 0.5 and
2 keV and 89 sources between 2 and 10 keV, at a flux limit of $1.3\times
10^{-15}$ and $3.4\times 10^{-15}$ erg cm$^{-2}$ s$^{-1}$, respectively;
68 of them  were detected in both energy bands.   In order to evaluate
the flux of  our sources, we assumed a template  AGN spectrum, i.e.  a
power--law  with photon--index  $\Gamma$=1.75 and  an  hydrogen column
density N$_{\rm  H}$ of $1.28\times  10^{21}$ cm$^{-2}$, corresponding
to the total galactic column density.
\vspace{-0.25 truecm}

\subsection{Log$N$--Log$S$ distribution}

In Fig.\ref{logNlogS} we show the cumulative log$N$--log$S$ distributions for the sources detected
in the two energy ranges.  For comparison, we have superimposed to our data the lower and upper
limits of the log$N$--log$S$ measured by \citet{Baldi2002} for a survey at high galactic latitude
($|b|>27^{\circ}$):  they obtained the upper limit log$N$--log$S$ by applying the same detection
threshold ($P_{\rm th}=2\times10^{-4}$) but a larger extraction radius, while the lower limit
log$N$--log$S$ was obtained with the same extraction radius but a more constraining threshold value
($P_{\rm th}=2\cdot10^{-5}$).  Moreover, in the same figure we have also reported the log$N$--log$S$
distributions, as well as the 90 \% confidence limits, measured by {\em CHANDRA} in the galactic
plane \citep{Ebisawa2005}.

In the soft energy band, the log$N$--log$S$ distribution of our sources is well above the high--latitude upper limit, expecially at low X--ray fluxes. Even if the galactic column density represents an overestimate for the stellar population of our sample, we have checked that not all of such an excess can be ascribable to
overcorrection for interstellar absorption arising from the use of the total galactic N$_{\rm H}$ value. We note also that $\sim$ 60 \% of the soft sources were not detected in the hard energy band.  In the soft band, the galactic plane log$N$--log$S$ distribution (the red points) is much lower than the one at high latitudes, since a significant fraction of extra--galactic sources is not detected.  Moreover, the same
log$N$--log$S$ is also lower than the difference between our data and the distribution limits at high latitudes (the blue lines).  Since \citet{Ebisawa2005}  find that most of their soft sources are nearby X--ray active stars, it is possible that our excess over their distribution is due to additional, more distant galactic sources, which are missed looking at $b\sim0^{\circ}$ but can be detected just outside the galactic plane.

In the hard energy band the distribution of our sources is in good agreement with both the high
latitude and the galactic plane ones measured by \XMM, {\em CHANDRA} and {\em ASCA}
\citep{Hands2004,Ebisawa2005}.  At energies $>$ 2 keV we expect the galactic absorption to be
negligible so that the extragalactic sources dominate the log$N$--log$S$ distribution at all
galactic latitudes, with just a small contribution of the {\em softer} galactic sources.
\vspace{-0.25 truecm}

\section{Search for optical counterparts}
In order to identify our serendipitous X--ray sources, we cross--correlated their positions with two
optical catalogues, namely
\begin{itemize}
\item the version 2.3 of the {\em Guide Star Catalogue} ({\em GSC}), not yet published, with
limiting magnitudes $B_{J} \sim 23$ and $F \sim 22$, photometric accuracy of $\sim$ 0.25 mag for $B_{J}$ and $\sim$ 0.2 mag for $F$, and position errors $<$ 0.5'' \citep{Chieregato2005}.
\item the {\em United States Naval Observatory} ({\em USNO}) catalogue \citep{Monet2003}, with
limiting magnitudes $V\sim21$, 0.2'' astrometric accuracy and $\sim0.3$ mag photometric accuracy.
\end{itemize}
\vspace{-0.25 truecm}

6 of our X--ray sources have a single bright, almost coincident, optical counterpart.  Since the
position error is much lower at optical wavelength ($\sim$ 0.5'') than for X--ray ($\sim$ 5''), we
used the optical positions to estimate the correction to be applied to the X--ray coordinates.
This turns out to be 1.83'' in RA and 1.44'' in DEC, for a total of 2.33''.

The search for optical counterparts was performed selecting candidates at $<$5'' from the corrected
position.  In such a way, we found at least one optical candidate counterpart for half of our
sources, namely 95 of the 196 sources.  Indeed, we found a total of 142 candidate optical
counterparts, since for 28 of the 95 X--ray sources we found more than one optical source within the
rather conservative 5'' radius error--circle.  It is not surprising that half of the detected X--ray
sources lack any optical counterpart:  in view of the length of our X--ray exposure, the expected
limiting magnitude of the possible counterpart is $V\simeq25$, much lower than the limiting
magnitude of the available catalogues.  Therefore, the identification of our fainter sources needs
ad hoc optical observations which are carried out at ESO.

The above results suggest that we cannot ignore the possible foreground contamination, which could
affect our cross--correlation.  The probability of chance coincidence between a X--ray and an
optical source is given by $P=1-e^{-\pi r^{2} \mu}$, where $r$ is the X--ray error--circle radius
and $\mu$ is the surface density of the optical sources \citep{Severgnini2004}.  In our case, within
the 15 arcmin radius imaged area the GSC catalogue provides a total of $\sim$16000 sources,
corresponding to a surface density $\mu\sim6.4\times10^{-3}$ sources arcsec$^{-2}$.  Since the
X--ray error--circle is 5 arcsec, we estimated that $P\simeq0.4$.  Therefore up to 40 \% of the
selected counterparts could be spurious candidates, in rough agreement with the number of X--ray
sources with multiple counterparts.
\vspace{-0.25 truecm}

\section{Bright source analysis}\label{bright}

\begin{figure}[!h]
\includegraphics[width=8.75cm]{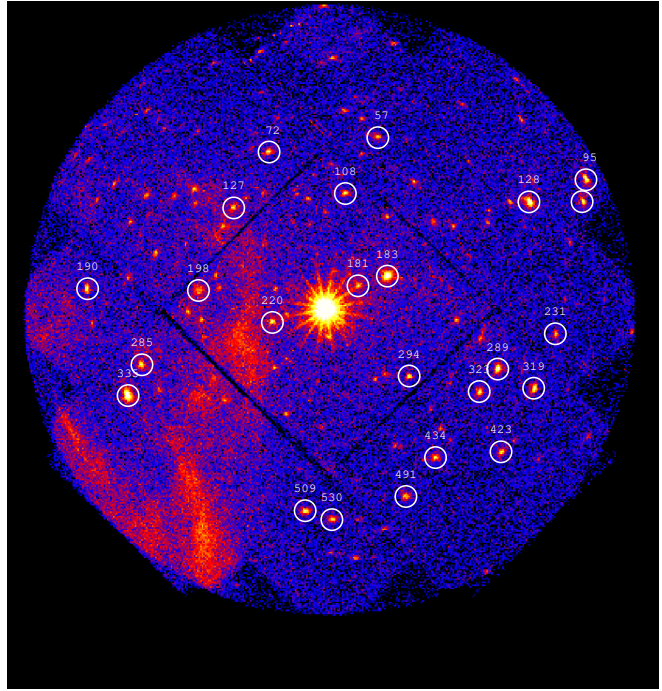}
\vspace{-0.5 truecm}
\begin{normalsize}
\caption{Image of the sky distribution of the 24 brightest sources, in the energy range
0.3--8 keV.}\label{bs}
\end{normalsize}
\end{figure}

Waiting for the optical data which will allow to characterise our sources on the basis of their
$f_{X}/f_{opt}$ ratio, we focused on the X--ray analysis of the brightest sources.  Since we
estimated that at least 500 counts are needed to discriminate thermal spectra from non--thermal
ones, we selected sources totalling $>$ 500 counts.  Out of our 196 sources, 24 satisfy this
requirement (Fig.~\ref{bs}).

We accumulated the source spectra by selecting only events with PATTERN=0--12 and generated {\em ad
hoc} response matrices and ancillary files using the {\em SAS} tasks {\em rmfgen} and {\em arfgen}.
Before spectral fitting, all spectra were binned with a minimum of 30 counts per bin in order to be
able to apply the $\chi^2$ minimization technique.  In this process, the background count rate was
rescaled using the ratio of the source and background areas.  Then we fitted the source spectra with
four spectral models:  {\em power--law}, {\em bremsstrahlung}, {\em black--body} and {\em mekal}
\footnote{{\em power--low}, {\em bremsstrahlung}, {\em black--body} and {\em mekal} are respectively {\tt pow},
{\tt bremss}, {\tt bbody} and {\tt mekal} in {\em XSPEC}} (i.e.  a bremsstrahlung model which includes also the element abundances); in all cases we included also the
absorption by the interstellar medium, leaving it as a free parameter.  For each emission model, we
calculated the 90 \% confidence level error on both the hydrogen column density and the
temperature/photon--index.  In this way we found that 13 sources were best fitted by a {\em
power--law} model, 2 by a {\em bremsstrahlung} model and 2 by a {\em mekal} model
(Tab.~\ref{sources}).  For 6 of the 7 remaining sources, at least two different models provided an
acceptable fit with a comparable $\chi_{\nu}^{2}$; finally, for source \#127 all the considered
models gave unacceptable results.

The spectral parameters were used to compute the sources' X--ray flux values, to be compared to the
optical ones in the framework of the $f_{X}/f_{opt}$ identification tool:  for the 23 sources with
at least one best--fit model we computed the X--ray flux based on the best--fit values, while for
source \# 127 we assumed a power--law spectrum with photon--index $\Gamma$ = 1.75 and a galactic
hydrogen column density.  On the optical side, we considered all the candidate
counterparts found within 5'' radius X--ray error circles. In order to minimize the effect of the interstellar extinction, we used the $F$ magnitude to calculate the source flux;  for the X--ray sources with no counterpart, we used $F$=22 as the optical upper limit.

On the basis of both the spectral fits (N$_{\rm H}$ and best--fit models) and the X--ray--to--optical flux ratios of the possible counterparts, we can propose a firm classification only for 7 sources, i.e.  6 AGNs and 1 star.  For 6 additional sources, the suggested classification (i.e.  4 AGNs and 2 stars) is affected by the best--fit value of the interstellar absorption, which is too low (for AGNs) or too high (for stars) in comparison with the galactic N$_{\rm H}$ (1.28$\times 10^{21}$ cm$^{-2}$).  In view of the large
errors on the N$_{\rm H}$ best--fit values, however, we accept the proposed identification.

\begin{landscape}
\begin{table}
\caption{Main characteristics of the 24 brightest sources. The sources are sorted by decreasing count number.}\label{sources}
\begin{tabular}{cccccccccccc} \hline
(1)	& (2)		& (3)		& (4)	& (5)		& (6)			& (7)			& (8)			& (9)		& (10)			& (11)			& (12)		\\
SRC	& RA (J2000)	& DEC (J2000)	& cts	& Model		& N$_{\rm H}$		& $\Gamma$/kT		& $\chi_{\nu}^{2}$	& D$_{XO}$	& $F$			& $\frac{f_{X}}{f_{F}}$	& SUGGESTED		\\
	& ($h m s$)	& ($^{\circ}$ ' '')	&	&		& (10$^{21}$ cm$^{-2}$)	& (-/keV)		&			& (arcsec)	& (mag)			& (log$_{10}$)		&	CLASS	\\ \hline
335	& 12 11 01.37	& -52 30 30.6	& 4229	& wabs(pow)	& 1.8$^{+0.4}_{-0.4}$	& 1.95$^{+0.14}_{-0.13}$	& 1.03			& 1.3, 3.1, 3.1, 3.5	& 20.36, 20.34, 19.35, 19.46	& 1.39, 1.38, 0.99, 1.03	& AGN		\\ \hline
183	& 12 09 41.88	& -52 24 56.9	& 3848	& 
wabs(pow)	& 0.7$^{+0.4}_{-0.4}$	& 1.90$^{+0.12}_{-0.13}$	& 1.17			& 1.5 		& 20.00			& 0.90			& ?		\\ \hline
128	& 12 06 58.63	& -52 21 26.1	& 2388	& wabs(mekal)	& 1.3$^{+0.7}_{-0.6}$	& 0.61$^{+0.03}_{-0.03}$	& 1.42			& 1.8		& 10.53			& -3.20			& STAR		\\ \hline
289	& 12 09 06.02	& -52 29 16.8	& 1499	& wabs(pow)	& 1.3$^{+0.5}_{-0.5}$	& 2.00$^{+0.16}_{-0.17}$	& 0.77			& 1.2		& 19.83			& 0.57			& AGN		\\ \hline
95	& 12 06 41.23	& -52 20 25.1	& 1332	& wabs(pow)	& 0.8$^{+0.6}_{-0.6}$	& 2.03$^{+0.22}_{-0.22}$	& 1.47			& -		& $>$22			& $>$1.49		& AGN (?)	\\ \hline
319	& 12 08 56.96	& -52 30 10.1	& 1109	& wabs(brem)	& 2.1$^{+2.0}_{-1.2}$	& 0.33$^{+0.16}_{-0.13}$	& 1.40			& 1.8		& 13.82			& -2.21			& STAR (?)	\\ \hline
509	& 12 10 07.06	& -52 35 54.2	& 1056	& wabs(pow)	& 0.8$^{+0.5}_{-0.6}$	& 1.78$^{+0.17}_{-0.19}$	& 1.25			& 1.2		& 19.06			& 0.29			& AGN (?)	\\ 
\hline
323	& 12 09 13.68	& -52 30 19.8	& 1000	& wabs(pow)	& 0.5$^{+0.6}_{-0.5}$	& 1.84$^{+0.20}_{-0.28}$	& 0.91			& -		& $>$22			& $>$1.26		& ?		\\
 \hline
190	& 12 11 13.66	& -52 25 30.7	& 984	& wabs(brem)	& 3.1$^{+2.4}_{-1.3}$	& 0.30$^{+0.13}_{-0.11}$	& 2.35			& 2.9		& 14.22			& -2.02			& ?		\\
	& 		& 		& 	& wabs(bbody)	& 1.5$^{+2.1}_{-0.5}$	& 0.17$^{+0.04}_{-0.03}$	& 2.39			& 2.9		& 14.22			& -2.02			& ?		\\ \hline
125	& 12 08 42.36	& -52 21 26.6	& 927	& wabs(pow)	& 0.1$^{+0.6}_{-0.1}$	& 1.75$^{+0.20}_{-0.22}$	& 1.01			& 3.1		& 18.9			& 0.14			& ?		\\ 
\hline
530	& 12 09 58.87	& -52 36 19.4	& 789	& wabs(pow)	& 1.6$^{+0.9}_{-0.9}$	& 1.80$^{+0.24}_{-0.24}$	& 0.93			& 2.9, 3.3		& 17.26, 16.95		& -0.52, -0.64		& AGN		\\ 
\hline
220	& 12 10 17.09	& -52 27 06.5	& 769	& wabs(brem)	& 1.8$^{+1.4}_{-1.0}$	& 0.40$^{+0.13}_{-0.13}$	& 1.01			& 3.7		& 15.42			& -1.81			& STAR (?)	\\ 
\hline
108	& 12 09 54.74	& -52 21 05.4	& 759	& wabs(pow)	& 1.1$^{+0.9}_{-1.0}$	& 2.02$^{+0.42}_{-0.29}$	& 0.82			& 2.4, 4.8		& 17.27, 18.91		& -0.90, -0.24		& AGN (?)	\\ 
\hline
72	& 12 10 18.05	& -52 19 09.1	& 746	& wabs(brem)	& 3.3$^{+2.8}_{-1.5}$	& 0.27$^{+0.12}_{-0.11}$	& 1.46			& 0.7		& 16.56			& -1.40			& ?		\\
	& 		& 		& 	& wabs(bbody)	& 1.5$^{+2.3}_{-0.6}$	& 0.17$^{+0.03}_{-0.05}$	& 1.45			& 0.7		& 16.56			& -1.39			& ?		\\ 
\hline
285	& 12 10 57.10	& -52 29 04.6	& 738	& wabs(mekal)	& 6.9$^{+3.4}_{-2.7}$	& 4.33$^{+3.84}_{-1.56}$	& 1.03			& 1.6, 2.8, 3.7	& 19.83, 20.43, 18.67	& +0.58, +0.82, +0.11	& ?		\\
	&		&		&	& wabs(pow)	& 7.2$^{+3.7}_{-2.3}$	& 1.95$^{+0.40}_{-0.29}$	& 1.15			& 1.6, 2.8, 3.7	& 19.83, 20.43, 18.67	& +0.59, +0.83, +0.12	& ?		\\
	&		& 		&	& wabs(bremss)	& 5.8$^{+2.7}_{-1.9}$	& 5.35$^{+5.45}_{-1.98}$	& 1.15			& 1.6, 2.8, 3.7	& 19.83, 20.43, 18.67	& +0.55, +0.79, +0.09	& ?		\\ 
\hline
434	& 12 09 27.05	& -52 33 25.2	& 736	& wabs(pow)	& 1.4$^{+0.8}_{-1.0}$	& 2.39$^{+0.40}_{-0.31}$	& 1.71			& -		& $>$22			& $>$1.11		& AGN (?)	\\ 
\hline
491	& 12 09 36.12	& -52 35 14.3	& 674	& wabs(pow)	& 1.4$^{+1.0}_{-1.1}$	& 1.93$^{+0.37}_{-0.29}$	& 1.09			& -		& $>$22			& $>$1.22		& AGN		\\
\hline
423	& 12 09 06.94	& -52 33 08.6	& 671	& wabs(pow)	& 2.1$^{+1.2}_{-1.1}$	& 2.18$^{+0.30}_{-0.35}$	& 0.54			& 1.2		& 19.72			& +0.26			& AGN		\\
\hline
181	& 12 09 50.88	& -52 25 24.2	& 669	& wabs(pow)	& 0.8$^{+0.9}_{-0.8}$	& 2.15$^{+0.25}_{-0.28}$	& 0.91			& -		& $>$22			& $>$0.90		& ?		\\
	&		&		&	& wabs(bremss)	& 0.0$^{+0.6}_{-0.0}$	& 3.20$^{+1.81}_{-1.20}$	& 0.97			& -		& $>$22			& $>$0.87		& ?		\\
	&		&		&	& wabs(mekal)	& 0.0$^{+0.3}_{-0.0}$	& 4.02$^{+1.72}_{-1.01}$	& 0.99			& -		& $>$22			& $>$0.91		& ?		\\
\hline
294	& 12 09 35.18	& -52 29 36.6	& 650	& wabs(bremss)	& 2.3$^{+2.0}_{-1.3}$	& 3.13$^{+2.72}_{-1.29}$	& 0.61			& -		& $>$22			& $>$0.90		& ?		\\
	&		&		&	& wabs(pow)	& 3.7$^{+2.8}_{-1.6}$	& 2.26$^{+0.59}_{-0.39}$	& 0.65			& -		& $>$22			& $>$0.93		& ?		\\ \hline
127	& 12 10 28.87	& -52 21 45.7	& 560	& ?		& -			& -			& -			& 1.3		& 14.93			& -1.80			& ?		\\ \hline
198	& 12 10 39.72	& -52 25 36.8	& 548	& wabs(mekal)	& 3.8$^{+1.6}_{-1.7}$	& 0.51$^{+0.09}_{-0.15}$	& 1.68			& 1.6		& 11.77			& -3.39			& ?		\\
\hline
57	& 12 09 44.83	& -52 18 26.8	& 532	& wabs(pow)	& 1.7$^{+1.2}_{-1.3}$	& 1.92$^{+0.50}_{-0.32}$	& 1.00			& -		& $>$22		& $>$+1.00		& AGN		\\
\hline
231	& 12 08 50.40	& -52 27 37.4	& 490	& wabs(bremss)	& 2.4$^{+2.2}_{-1.4}$	& 0.46$^{+0.23}_{-0.18}$	& 1.88			& 1.3		& 16.25			& -1.54			& ?		\\
	&		&		&	& wabs(bbody)	& 0.6$^{+2.1}_{-0.6}$	& 0.23$^{+0.05}_{-0.06}$	& 1.92			& 1.3		& 16.25			& -1.54			& ?		\\ \hline
\end{tabular}

Key to Table - Col.(1):  Source ID number.  Col.(2) and (3):  source celestial coordinates.
Col.(4):  source total counts (in the energy range 0.3--8 keV).  Col.(5):  best--fit emission
model(s); the symbol `?'  indicates that none of the tested single--component models provided an
acceptable fit.  Col.(6):  best--fit hydrogen column density, with the relevant 90 \% confidence
level error for two interesting parameters ($\Delta \chi^2=4.61$).  Col.(7):  best--fit
photon--index or plasma temperature, in the case of either a power--law or a thermal emission model,
respectively; also the relevant 90 \% confidence level error for two interesting parameters ($\Delta
\chi^2=4.61$) is reported.  Col.(8):  best--fit reduced chi--squared.  Col.(9):  projected sky
distance, from the X--ray position, of the candidate optical counterpart (if any).  Col(10):  $F$
magnitude of the optical candidate counterpart; we consider $F>$22 if no candidate counterpart is
found within a 5'' radius X--ray error circle.  Col.(11):  logarithmic values of the
X--ray--to--optical flux ratio; the optical flux is based on the $F$ magnitude; the X--ray flux is
based on the best--fit model or, when no model is acceptable, on a power--law spectrum with
photon--index $\Gamma$ = 1.75 and hydrogen column density N$_{\rm H}=1.28\times 10^{21}$ cm$^{-2}$,
corresponding to the total galactic column density.  Col.(12):  proposed source classification; the
symbol `?'  after it means that the reported identification is uncertain, due the N$_{\rm H}$ value;
the only symbol `?'  indicates that no classification can be suggested.
\end{table}
\end{landscape}

\begin{figure}[htb]
\includegraphics[width=5.5cm,angle=-90]{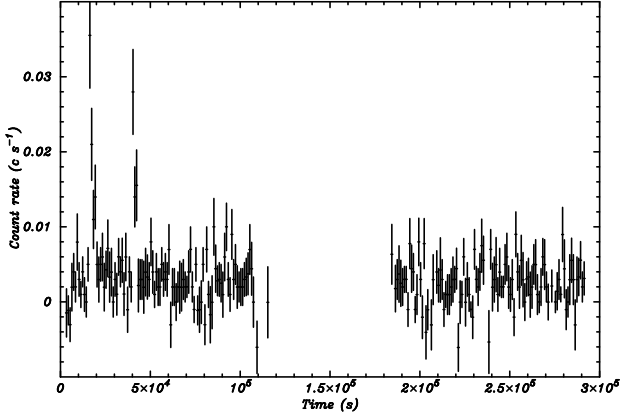}
\begin{normalsize}
\caption{Light--curve of source \#72, with a 1 ksec time binning.}\label{lc72}
\end{normalsize}
\end{figure}

4 additional sources (\# 190, 72, 198 and 231) are characterized both by a low temperature thermal
spectrum and by a low X--ray--to--optical flux ratio, therefore it is probable that they are stars.
Unfortunately they have a high N$_{\rm H}$ value and, in 3 cases, also the emission model is
uncertain, therefore the star identification can not be firmly established.  For source \# 72 this
classification would be supported also by the observed light curve (Fig.~\ref{lc72}), which shows
large but short flares and a flux variability with time--scales of a few hundred seconds.

We note that single component fitting can induce further uncertainty on the N$_{\rm H}$ estimate. Indeed, stars do show two temperature spectra (actually coronal loop distributions) which, if fitted with a single temperature, would result in an overestimate of the N$_{\rm H}$ values. AGNs, on the other hand, often have additional soft components which, for a pure power--law fit, would yield too low N$_{\rm H}$ values. In view of the above uncertainties, we underline that the  source classification  proposed  in Tab.~\ref{sources} is only tentative.
\vspace{-0.5 truecm}

\vspace{-0.5 truecm}

\begin{figure}[htb]
\includegraphics[width=5.5cm,angle=-90]{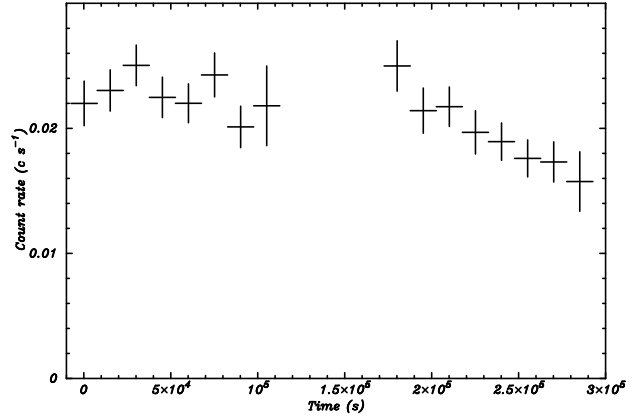}
\begin{normalsize}
\caption{Light--curve of source \#183, with a 15 ksec time binning.}\label{lc183}
\end{normalsize}
\vspace{-0.5 truecm}
\end{figure}

Only the low N$_{\rm H}$ value prevents to classify as AGNs 3 other sources (\# 183, 323 and 125),
which are best fitted by a power--law spectrum with photon index $\simeq$2 and have a rather high
X--ray--to--optical flux ratio.  The smooth variability observed for source \# 183, with a
time--scale of $\sim 10^{4}$ s (Fig.~\ref{lc183}), would also support an AGN identification
\footnote{Even if, given the maximum estimated luminosity (L$_{X}$ $\sim$ 10$^{32}$ erg s$^{-1}$) of
a possible galactic counterpart (6.4 kpc), this source could be also a quiescent LMXRB or CV}.

For 3 sources with hard spectrum (\# 285, 181 and 294) it is not possible to distinguish between a
power--law and a high temperature thermal emission model:  with all models sources \# 285 and 294
show a high N$_{\rm H}$ value, therefore they are probably extragalactic objects (either AGNs or
clusters of galaxies).  On the other hand, in all cases source \# 181 has a very low best--fit value
of N$_{\rm H}$, therefore it should be a galactic object, even if its nature can not be established.

Finally, source \# 127 has a very unusual spectrum and it will be discussed in detail in
Sec.~\ref{127}.

On the basis of the above results, we conclude that 8 sources over 23 (i.e.  $\sim$ 35 \%) could
belong to the Galaxy.  Such a percentage is in agreement with the results obtained by previous {\em
ROSAT} surveys which showed that the stellar content decreases from $\sim$85\% to $\sim$30\% moving
from the galactic plane to high galactic latitudes \citep{Motch1997,Zickgraf2003}.

\section{Source \#127}\label{127}

\begin{figure}[htb]
\centerline{%
\includegraphics[height=8.5cm,angle=-90]{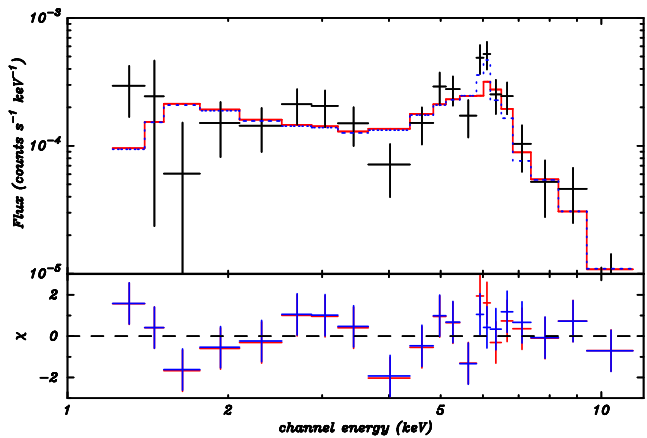}}
\begin{normalsize}
\caption{{\em Top}:  comparison of the unbinned spectrum of source \#127 with the
best--fit model, in the case of both redshift fixed at $z$=0.032 (red solid line) and of
best--fit value $z$=0.057 (blue dotted line).  {\em Bottom}:  data - model residuals (in
$\sigma$) for the two above models.}\label{spectrum}
\end{normalsize}
\end{figure}

The X--ray analysis yields 560 counts in the energy band 0.3--8 keV, with a signal--to--noise ratio
of 14.64; its count rate in the total energy band is 2.03$\times$10$^{-3}$ cts s$^{-1}$.  The source
spectrum cannot be described by a standard single--component emission model (Fig.~\ref{spectrum}):
it is very hard and highly absorbed; moreover, it is also characterized by a feature at $\sim$6 keV,
ascribable to Fe emission line.

After the astrometric correction, the resulting X--ray position is $\alpha_{J2000}$=12$^h$ 10$^m$
28.87$^s$, $\delta_{J2000}$=-52$^{\circ}$ 21' 45.7''.  Searching the {\em NED} ({\em Nasa/Ipac
Extragalactic Database}) we found the spiral galaxy ESO 217-G29, located at 1.28" from the X--ray
source position.  The magnitudes of ESO 217--G29 are $B_{J}=16.74$ and $F=14.93$ and its redshift is
$z=0.032$ \citep{Visvanathan1992}.  These parameters, together with the X--ray spectrum and the
estimated X--ray--to--optical flux ratio, suggest that source \#127 could be an AGN.

The source is located within the region of diffuse emission (Fig.  \ref{dr}), so its spectrum at low
energies (E$<$1keV) is polluted by the supernova remnant.  Thus we fit the source spectrum only
above 1.2 keV.  According to the AGN unification model \citep{Antonucci1993,Mushotzky1993}, the
source spectrum $S$ has been described by the model

\vspace{+0.25 cm}
$S=A_{G}[A_{SP}(R_{W})+A_{T}(PL+R_{C}+GL)]$\footnote{{\tt wabs*(zwabs*powerlaw + zwabs*(powerlaw + pexrav + zgauss))} in {\em XSPEC}}
\vspace{+0.25 cm}

where $A_{G}$ is the galactic absorption (1.28$\times$10$^{21}$ cm$^{-2}$), $A_{SP}$ is the
absorption related to the galaxy hosting the AGN, $R_{W}$ is the warm and optically thin reflection
component, $A_{T}$ is the absorption acting on the nuclear emission associated to the torus of dust
around the AGN nucleus, $PL$ is the primary power--law modeling the nuclear component, $R_{C}$ is
the cold and optically thick reflection component and $GL$ is the Gaussian component that models the
Fe line at 6.4 keV.  For the $A_{SP}$, $A_{T}$, $R_{C}$ and $GL$ components the redshift value is
fixed at $z$=0.032 \citep{Visvanathan1992}.

\begin{table}[htbp]
\caption{Best--fit parameters for source \# 127, for the optical redshift $z$=0.032 and for its best--fit value $z$=0.057.}\label{fit}
\begin{tabular}{cccc} \hline
Component 	& Parameter		& $z$=0.032 (fix)			& $z$=0.057			\\ \hline
$A_{SP}$		& N$_{\rm H1}^{a}$	& 2.26$_{-1.10}^{+1.42}$		& 2.39$_{-1.14}^{+0.81}$		\\
$R_{W}$		& $\Gamma$		& 1.9 (fixed)			& 1.9 (fixed)			\\
		& Flux @ 1 keV$^{b}$	& 7.53$_{-2.78}^{+3.43}$		& 7.53$_{-2.63}^{+3.52}$		\\
$A_{T}$		& N$_{\rm H2}^{a}$	& 75.82$_{-19.10}^{+25.02}$	& 82.35$_{-24.23}^{+18.69}$	\\
$PL$		& $\Gamma$		& 1.9 (fixed)			& 1.9 (fixed)			\\
		& Flux @ 1 keV$^{c}$	& 1.93$_{-0.80}^{+1.37}$		& 1.98$_{-0.71}^{+1.51}$		\\
$R_{C}$		& $\Gamma$		& 1.9 (fixed)			& 1.9 (fixed)			\\
		& Flux @ 1 keV$^{c}$	& 1.93$_{-0.80}^{+1.37}$		& 1.98$_{-0.71}^{+1.51}$		\\
$GL$		& $E_{line}$(keV)		& 6.4 (fixed)			& 6.4 (fixed)			\\
		& $I_{line}^{d}$		& 1.26$_{-1.26}^{+1.81}$		& 2.33$_{-1.46}^{+2.42}$		\\
		& EQW (eV)		& 185$_{-185}^{+265}$		& 311$_{-196}^{+322}$		\\ \hline
d.o.f.		&			& 32				& 31				\\
$\chi^{2}_{\nu}$	&			& 1.143				& 1.035				\\ \hline
\end{tabular}
\begin{small}
\\
$^{a}$ $10^{22}$ cm$^{-2}$

$^{b}$ $10^{-6}$ ph cm$^{-2}$ s$^{-1}$ keV$^{-1}$

$^{c}$ $10^{-4}$ ph cm$^{-2}$ s$^{-1}$ keV$^{-1}$

$^{d}$ $10^{-6}$ ph cm$^{-2}$ s$^{-1}$
\end{small}
\end{table}
\vspace{-0.5 truecm}

The best--fit parameters, listed in Tab.~\ref{fit}, provide an acceptable fit, yielding
$\chi^{2}_{\nu}$=1.143 with 32 d.o.f.; the value of N$_{\rm H2}$ implies that the torus around the
AGN is Compton--thin.  However, this model does not describe satisfactorily the prominent Fe line,
since it assigns an energy of 6.2 keV to the line centroid (red solid line in Fig.~\ref{spectrum}),
while in the accumulated spectrum the line is centered around 6.0 keV; moreover, the line
significance is marginal.
\vspace{-0.5 truecm}

Leaving the $z$ value as a free parameter, we obtain a better fit ($\chi^{2}_{\nu}$=1.035 with 31
d.o.f) for $z=0.057_{-0.016}^{+0.009}$, quite different, although consistent at 2 $\sigma$ level,
with the optical value; moreover, the line is significant at 90 \% confidence level.  Using the
F--test, the improvement with respect to the previous fit based on the optical redshift is
significant at 95\% confidence level.  In Tab.~\ref{fit} we report the best--fit parameters of both
fits.  As a further check, we applied also the Cash statistics to the {\em XSPEC} fit and we
obtained the same results:  $z$=0.057 and a normalization of 2.26$_{-1.34}^{+2.14}$ for the iron
line.  If we compare the source spectrum with the best--fit model (blue dotted line in
Fig.~\ref{spectrum}), we note that the Fe line is modelled more accurately and is centered around
6.0 keV.

The discrepancy between the X--ray and optical redshift values could be explained by the
relativistic broadening of the Fe line.  Recently, the rest--frame spectra of several sources
detected in the \XMM~survey of the Lockman hole showed a relativistically broadened iron line
\citep{Streblyanska2004}.  Owing to the Compton--thin nature of our source, it is possible that we
are observing the same phenomenology.  This would explain why the best--fit redshift overcomes the
cosmological one.  We investigated this possibility by modelling the Fe line with a relativistic
line ($RL$) from an accretion disc.  To this aim we replaced the Gaussian component of our model
with either a {\em laor} \citep{Laor1991} or a {\em diskline} \citep{Fabian1989} component
\footnote{respectively, {\tt laor} and {\tt diskline} in {\em XSPEC}}, leaving $z$=0.032 for the
other components.  We fixed the emissivity index $\beta$ to 3 and to -2 for the {\em laor} and the
{\em diskline} case, respectively; moreover, in both cases we fixed the line energy to 6.4 keV and
the disc inclination angle $i$ to 30$^\circ$, which is near the best--fit value found by
\citet{Streblyanska2004}.

\begin{figure}[htb]
\centerline{%
\includegraphics[height=8.5cm,angle=-90]{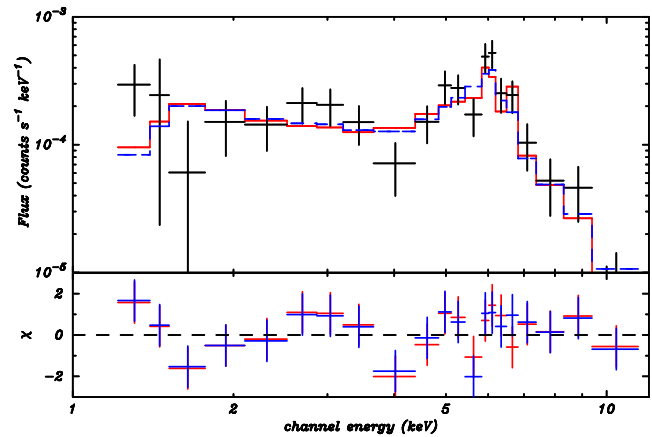}}
\begin{normalsize}
\caption{{\em Top}:  comparison of the unbinned spectrum of source \#127 with the
best--fit model and $z$=0.032, in the case of both a {\em laor} (red solid line) and a {\em diskline}
(blue dotted line) model for the Fe line.  {\em Bottom}:  data - model residuals (in
$\sigma$) for the two above models.}\label{secondspectrum}
\end{normalsize}
\end{figure}

In both cases the best--fit model traces rather well the Fe line (Fig.~\ref{secondspectrum}) and
provides an acceptable fit, yielding $\chi^{2}_{\nu}$=1.034 and 1.087 for the {\em laor} and the
{\em diskline} component, respectively.  For both models we find that the relativistic component is
significant at 90 \% confidence level. However, the disc inner and outer radii values are too small (i.e. a few $R_{g}$) and their difference is not significant. Moreover, 
only for the {\em laor} component the line EQW is
comparable to the value of $\sim$0.4 keV found by \citet{Streblyanska2004}, while it is
significantly larger ($\sim$ 1 keV) for the {\em diskline}.  Since these parameters are
affected by large errors, due the low count statistics, we conclude that the iron line
position can be reconciled with the redshift of the proposed optical counterpart ESO 217-G29.
\vspace{-0.5 truecm}

The 2--10 keV unabsorbed flux of the primary nuclear component is 5.79$_{-2.40}^{+4.11}\times
10^{-13}$ erg cm$^{-2}$ s$^{-1}$ (calculated with {\em XSPEC}).  Such a flux value, together with
the optical magnitude, implies that $f_{X}/f_{opt}$=0.41, i.e.  well within the AGN range
\citep{Krautter1999}.  The X--ray luminosity of the source in the 2--10 keV energy band, corrected
by the absorption and with the redshift at 0.032, is 2.59$_{-1.07}^{+1.84}\times 10^{42}$ erg
s$^{-1}$, corresponding to a low luminosity Seyfert galaxy.

Thus, the X--ray spectrum, together with the best--fit value of {\em N$_{\rm H2}$} and the nature of
the optical candidate counterpart led us to propose that source \#127 could be a new,
low--luminosity Seyfert--2 galaxy discovered serendipitously in our field.

\section{Summary and conclusions}

The longest \XMM~observation at low galactic latitude yielded a sample of 135 sources between 0.5
and 2 keV and of 89 sources between 2 and 10 keV, with limiting fluxes of $1.3\times 10^{-15}$ and
$3.4\times 10^{-15}$ erg cm$^{-2}$ s$^{-1}$, respectively.  The log$N$--log$S$ distribution of the
hard sources is comparable to that measured at high galactic latitudes, thus suggesting that it is
dominated by extragalactic sources.  On the other hand, at low fluxes the distribution of the soft
sources shows an excess above both the Galactic Plane and the high--latitude distributions:  we
consider this result as a strong indication that we observed a sample of both galactic and
extragalactic sources.

We analysed the 24 brightest sources and proposed an identification for $\sim$ 80 \% of them.
Moreover, the detailed spectral investigation of one unidentified source, characterized by a highly
absorbed spectrum and an evident Fe emission line, led us to classify it as a new Seyfert--2 galaxy.

The full X--ray characterization of all the sources, as well as their classification, based on ad
hoc optical observations, will be discussed in future papers.

\begin{acknowledgements}

We are grateful to K. Ebisawa for providing us the log$N$-log$S$ data of the {\em Chandra}
observation of the galactic plane.  We wish to thank the referee for his useful comments, which improved the presentation of our results. We also thank S.  Molendi and A.  Tiengo for their suggestions and stimulating discussions. This work is based on observations obtained with \XMM, an ESA science mission with instruments and contributions directly funded by ESA Member States and NASA.  The \XMM~data analysis is supported by the Italian Space Agency (ASI).  ADL acknowledges an ASI fellowship.  GN acknowledges a `G.  Petrocchi' fellowship of the Osio Sotto (BG) city council.  The Guide Star Catalog used in this work was produced at the Space Telescope Science Institute under U.S.  Government grant.  These data are based on photographic data obtained using the Oschin Schmidt Telescope on Palomar Mountain and the UK Schmidt Telescope.  This research has made use of the USNOFS Image and Catalogue Archive operated by the United States Naval Observatory at the Flagstaff
Station (http://www.nofs.navy.mil/data/fchpix/) and of the NASA/IPAC Extragalactic Database (NED) which is operated by the Jet Propulsion Laboratory, California Institute of Technology, under contract with the National Aeronautics and Space Administration.

\end{acknowledgements}
\vspace{+0.5 truecm}

\bibliographystyle{aa}
\bibliography{biblio}

\end{document}